**Magneto-optical evidence for single-crystal-like magnetic switching of epitaxial antiferromagnetic $LaFeO_3$ films**

A. Rieche,[1] W. Hoppe,[1] C. Körner,[1] A. D. Rata,[1] F. Weber,[2,3] J. B. G. Danziger, E. M. Vocks,[1] F. Wührl,[1] M. Bargheer,[2,3] W. Widdra,[1] G. Woltersdorf,[1] S. Ebbinghaus,[1] A. Herklotz,[1] K. Dörr[1]

[1]Institute of Physics, Martin-Luther-Universität Halle-Wittenberg, 06099 Halle, Germany

[2]Helmholtz-Zentrum Berlin für Materialien und Energie GmbH, Wilhelm-Conrad-Röntgen Campus, BESSY II, 12489 Berlin, Germany

[3]Institute of Physics and Astronomy, University of Potsdam, Karl–Liebknecht–Str. 24–25, 14476 Potsdam, Germany

**Abstract**

Strained epitaxial films of the antiferromagnetic orthoferrite $LaFeO_3$ offer a promising platform for antiferromagnetic spintronics, yet their magnetic switching behavior and domain structure have remained largely unexplored due to the small magnitude of the weak ferromagnetic moment. Here, we demonstrate that longitudinal magneto-optical Kerr effect (MOKE) measurements provide a sensitive and direct probe of magnetic switching and domain processes in coherently strained $LaFeO_3$ thin films grown on orthorhombic substrates. By employing $DyScO_3$(110), $GdScO_3$(110), and $NdGaO_3$(110) substrates, we achieve strain-controlled, largely twin-free growth and identify the orientation of the orthorhombic ***c***-axis through the presence or absence of a longitudinal MOKE signal. Compressively strained films exhibit large Kerr signals, rectangular hysteresis loops, and magnetic single-domain remanence over macroscopic areas. Tensile strain on orthorhombic substrates is associated with two competing structural effects on thin film orientation; in-plane magnetization has been identified in some films on $GdScO_3$(110) by MOKE. Angle-dependent MOKE hysteresis follows the Kondorsky model, indicating domain-wall-controlled switching analogous to bulk single crystals. Kerr microscopy reveals abrupt domain nucleation and rapid domain-wall motion, with defects acting as pinning centers and governing the coercive field. Our results establish MOKE as an efficient optical tool for identifying orthorhombic orientation, probing magnetic switching of coupled weak magnetization and Néel vectors, and accessing domain dynamics in $LaFeO_3$ films. This provides a foundation for strain-engineered orthoferrite thin films in antiferromagnetic spintronics and magnonics.

## Introduction

Recent strong interest in orthoferrites $RFeO_3$ (R = rare earth element or Y) arose in the search for antiferromagnetic materials promising faster spintronics (*1 - 3*). The large Néel temperature of up to 740 K and their fast spin dynamics make the orthoferrites interesting candidates for spintronics applications. However, all experiments with ultrafast optical excitation have been conducted on bulk crystals. Early interest in films had a different origin, namely the large x-ray magnetic linear dichroism (XMLD) resulting from the antiferromagnetically ordered Fe moments. This made possible the imaging of nanoscale antiferromagnetic domains for the first time in any antiferromagnetic film (*4*). Subsequent work studied domain correlations in thin-film $LaFeO_3$/ferromagnet bilayers and nanostructures employing x-ray absorption and x-ray photoelectron emission microscopy (PEEM) techniques. For examples, see Refs. *5*, *6*. Recently, the Spin Hall magnetoresistance of $LaFeO_3$ films has been investigated and related to switching the antiferromagnetic vector (***L***) by an applied magnetic field, together with the weak canted magnetization (***M***) (*7*, *8*).

$LaFeO_3$ as the starting member of the orthoferrite series $RFeO_3$ is an orthorhombic perovskite-type insulator with antiferromagnetic order of $Fe^{3+}$spins. The Dzyaloshinskii-Moriya (DM) interaction in conjunction with single-ion anisotropy causes a canting of Fe spins by an angle of about 0.53° in bulk crystals (*9*). Weak ferromagnetic magnetization (***M***) arises as result of the canting. In bulk crystals, ***M*** is aligned along the orthorhombic ***c***-axis (which is the one axis with in-phase rotations of oxygen octahedra, in the Glazer notion of $\boldsymbol{a^-b^-c^+}$ (*Pbnm*)). The orientation of Fe spins and the resulting antiferromagnetic vector is along the shorter of the remaining axes which is the orthorhombic ***a***-axis in this notation. The lattice structure with spin orientations is shown in Fig.1.

Thin epitaxial films in most early and some recent work have been grown on cubic $SrTiO_3$(001) substrates (*4 - 6*). The symmetry of the substrate allows for six possible orientations of the orthorhombic unit cell. Park et al. reported on the growth of coherently strained $LaFeO_3$ films on orthorhombic $GdScO_3$(110) substrates; their films have been thoroughly structurally characterized and appear to be orthorhombic twin-free (*10*). Kjaernes et al. grew coherently strained films on various orthorhombic scandate substrates with pseudocubic (111) orientation (*11*). They could identify an antiferromagnetic single-domain state in XMLD experiments. Recently, Polewczyk et al. systematically determined by XMLD the orientation of the Néel vector (***L***) in a series of epitaxially strained $LaFeO_3$ films on substrates spanning an average in-plane strain from -1.8% to +1.5% (*12*). In their experiments, epitaxial strain reorients the Néel vector from close to in-plane to out-of-plane with growing strain. (We note that this fascinating strain effect on the Néel vector in $LaFeO_3$ films could be surface- or volume-dominated, since the signal depth of XMLD is restricted.)

Ultrafast spin dynamics often requires optical detection. XMLD is suitable, but means rather large experimental efforts and limited measuring opportunities, since measurements usually require a synchrotron facility. As a versatile alternative available in many labs, the significant linear magnetooptical Kerr effect (MOKE) of $LaFeO_3$ can be used for magnetic characterization. Spectra of the magnetooptical Kerr effect of orthoferrite crystals have been reported by Kahn et al. (*13*). That work does not include $LaFeO_3$, but many other members of orthoferrites $RFeO_3$. According to our knowledge, there is yet just one publication on MOKE of $LaFeO_3$ films, for $LaFeO_3$/$DyScO_3$(110) by Alaria et al (*14*). For another orthoferrite, $YFeO_3$, Wang et al. reported on longitudinal MOKE of twin-free films on $NdGaO_3$(110), demonstrating unusually large (≥ 0.25 mm) magnetic domains (*15*).

Density functional theory was employed by Mao et al. to investigate the preferred orientation of the orthorhombic unit cell and the magnetic structure of $LaFeO_3$ in dependence on isotropic in-plane strain in films. (*16*) Their result predicts a structural reorientation between states of tensile strain (***c***-axis perpendicular to the film plane) and compressive strain (***c***-axis in the film plane) (Fig.1). An experimental investigation of the orthorhombic orientation of $EuFeO_3$ films with similar crystal structure as $LaFeO_3$ (*17*) agrees with this predicted strain effect. Additionally, for orthorhombic $GdScO_3$(110) substrates, a second and competing effect has been identified by Choquette et al. (*17*): the transfer of oxygen octahedral rotations from the substrate to the film which, in their $EuFeO_3$/$GdScO_3$(110) sample, aligns the ***c***-axes of film and substrate in-plane in spite of the large tensile strain. Hence, on orthorhombic substrates, the balance between both effects seems to decide about the orthorhombic orientation of films under tensile strain. This makes it highly desirable to find an experimental tool for routine identification of the magnetization axis ***M*** of $LaFeO_3$ and other $RFeO_3$ films.

The present work shows that coherently strained $LaFeO_3$ thin films grown on orthorhombic substrates can exhibit single-crystal-like magnetic switching behavior that is directly accessible by longitudinal magneto-optical Kerr effect measurements. For compressively strained films, the magnetization lies in the film plane and gives rise to large Kerr signals, rectangular hysteresis loops, and macroscopic single-domain remanence. Angle-dependent hysteresis measurements follow the Kondorsky model, indicating domain-wall-controlled switching analogous to bulk $LaFeO_3$ crystals. These results establish MOKE as a sensitive tool for identifying orthorhombic orientation, probing magnetic switching, and imaging domain processes in antiferromagnetic orthoferrite thin films. Magnetization hysteresis curves of a $LaFeO_3$ single crystal have been investigated as reference for the MOKE hysteresis curves of films, since published work only shows the magnetization hysteresis of other orthoferrites, for example $YFeO_3$ (*18*, *19*) and $TmFeO_3$ (*20*).

## Experiment

### Thin film growth

Epitaxial $LaFeO_3$ (LFO) films were grown on various substrates using pulsed laser deposition (PLD), in a vacuum chamber (*Surface GmbH*) equipped with an excimer laser operating at a wavelength of 248 nm. The stoichiometric LFO target required for this process was produced using standard solid-state synthesis. Orthorhombic substrates of $NdGaO_3$(110), $DyScO_3$(110) and $GdScO_3$(110) were chosen to support twin-free growth of the orthorhombic films. The LFO films were prepared at a laser fluence of 1 J $cm^{-2}$ and a repetition rate of 2 Hz. The growth temperature and oxygen background pressure were 700 °C and 0.2 mbar, respectively. After growth, the samples were cooled in 200 mbar of oxygen to ensure full oxidation of the films. Structural properties were characterized by high-resolution x-ray diffraction (XRD) in a Bruker

Advance D8 diffractometer setup equipped with monochromatic Cu $K\alpha_1$ radiation ($\lambda$ = 1.54056 Å).

Crystal growth

Rods for the crystal growth were obtained by hydrostatic pressing of polycrystalline $LaFeO_3$ (prepared by classical solid-state synthesis) followed by sintering in air at 1430 °C for 12 h. Crystal growth was carried out in a *Crystal Systems Corporation* optical floating zone furnace model FZ-T-10000-H-VPO-PC equipped with four 1000 W halogen lamps. A gas flow of 100 ml/min synthetic air was used, and the feed and seed rods were counter-rotated with 25 rpm and 30 rpm, respectively. A piece of the obtained crystal was oriented parallel to the pseudo-cubic (001) plane on a four-circle diffractometer, followed by rocking curve measurements on a Bragg-Brentano powder diffractometer to an accuracy better than ± 0.5°.

Magneto-optical Kerr effect (MOKE)

To study the magnetic properties of the LFO films, the magneto-optical Kerr effect is employed. When polarized light reflects from a magnetically ordered sample, the polarization state of the reflected light is altered due to the magneto-optical Kerr effect. Depending on the relative alignment of the sample, the magnetic order and the plane of incidence of the light, one distinguishes between polar, longitudinal, and transversal MOKE. In longitudinal geometry, the Kerr effect leads to a rotation of the linear polarization by the Kerr angle $\boldsymbol{\vartheta}_{Kerr}$, which is proportional to the magnetic order parameter (***M***). We utilize this effect as the magnetic order lies in the plane of the sample as well as the plane of incidence. The experimental scheme is the following: the light of a 405 nm laser diode gets s-polarised (perpendicular to the plane of incidence) and, after reflection, passes an analyser which is slightly rotated out of p-polarisation (parallel to the plane of incidence). The transmitted intensity is then detected by a single diode photodetector as a voltage $\boldsymbol{U}_{\mathbf{PD}}$. Reflection upon a magnetic sample rotates the light's polarisation proportional to ***M*** and thus changes $\boldsymbol{U}_{\mathbf{PD}}(\boldsymbol{M})$ accordingly. Using an electromagnet to apply an external magnetic field up to ±1.5 T allows to obtain hysteresis loops of our samples. Note that a 405 nm light source is used as $\boldsymbol{\vartheta}_{Kerr}$ for LFO is strongly wavelength dependent (*13*).

While this scheme allows for a quick characterization of the sample, the resulting $\boldsymbol{U}_{\mathbf{PD}}$ represents the spatial average of ***M*** across the laser spot size (≈ 0,5 mm²) and, thus, cannot resolve domain structures.

To achieve spatial resolution, a related experiment is conducted: Kerr microscopy. Here, a special white light source provides eight individually controllable LEDs which are coupled into the microscope using individual fibers arranged in a cross-shape at the entrance of the microscope. After passing a linear polarizer, the light from these fibers is imaged on the sample using an objective lens. Switching the individual illumination LEDs allows to set a preferential ***k***-vector of the light incident on the sample, which is necessary to achieve contrast due to Kerr rotation of the polarization. The reflected light passes a waveplate to compensate for ellipticity and an analyser, before it is imaged on a camera sensor. This allows to capture a full wide field image of the sample with diffraction-limited resolution, which is only slightly decreased due to the uneven illumination of the objective lens. An electromagnet allows to generate in-plane magnetic fields of up to ±1 T to influence the magnetization state of the sample and to capture a hysteresis loop.

Single-crystal magnetization measurement

Magnetic properties of single crystal LFO were investigated using a Quantum Design PPMS9 system. Hysteresis loops were taken at 300 K and 10 K, with magnetic DC field cycling between ±9 T. Low-field magnetization was also measured using a MPMS-SQUID magnetometer.

## Results

Crystal structure and strain state of films

The orthorhombic lattice parameters (at 300 K) of the used bulk materials are listed in Tab.1. The averaged in-plane lattice parameter of the (110)-oriented orthorhombic substrates is calculated as $\boldsymbol{a_{pc}}$ = ¼ [sqrt($\boldsymbol{a}^2$+$\boldsymbol{b}^2$)+ $\boldsymbol{c}$]. The pseudocubic bulk lattice parameter of $LaFeO_3$ is $\boldsymbol{a}_{\mathbf{bulk}}$ = (¼ $\boldsymbol{a\ b\ c}$)$^{1/3}$ = 3.930 Å. We note that even though $LaFeO_3$ is orthorhombic, its crystal distortion with respect to the related cubic structure is very small. Therefore, orthorhombic splitting of film peaks was not detectable by in-house x-ray diffraction. Pseudocubic lattice parameters (marked with a subscript "pc") have been used for estimating the average film strain $\varepsilon = 1 - \boldsymbol{c_{pc}}/\boldsymbol{a_{pc}}$, with the measured value of $\boldsymbol{c_{pc}}$ (the pseudocubic out-of-plane lattice parameter). The strain is negative (positive) for the case of in-plane compression (expansion). The volume expansion of the LFO pseudocubic unit cell is estimated as $\Delta\boldsymbol{V} = \boldsymbol{a_{pc}}^2\boldsymbol{c_{pc}} - \boldsymbol{a}^3_{\boldsymbol{bulk}}$.

For this study, coherently grown $LaFeO_3$ films on all substrates have been used as checked by x-ray reciprocal space mapping (Suppl. Figs.S1, S3, S5). Therefore, the averaged film in-plane lattice parameter ($\boldsymbol{a_{pc}}$) is assumed as identical to that of the substrate. The film out-of-plane lattice parameter ($\boldsymbol{c_{pc}}$) is derived from the position of film peaks or, in case of overlap with substrate peaks, from superlattice reflections.

The Supplement includes a sample table (Tab.S1) where the film thickness, the average strain ($\varepsilon$), the unit cell volume increase ($\Delta\boldsymbol{V}$) and the presence of in-plane magnetization (L-MOKE) of a few representative films are listed. On DSO(110), films with thicknesses from 6 nm to 70 nm have been grown. Films grew in a layer-by-layer mode for thicknesses < 30 nm, thicker films showed some roughness with small islands. The pseudocubic lattice parameter of LFO is about 0.4 % smaller than the averaged in-plane lattice parameter of DSO(110) substrates. However, we observed some films under weak compressive strain. The films can have a slightly increased volume of the unit cell (< 2.5 %, Tab.S1), depending on growth conditions. Usually undesirable, because imperfections of the grown crystal lattice like vacancies are responsible for the lattice expansion, this moderate volume expansion is frequently observed in perovskite-type oxide films (e. g., as investigated in Ref. *21*). Here, it can be utilized to grow films in various strain states on DSO(110). Films on the larger GSO(110) with misfit of about 0.7% grow with tensile strain in the film plane. Films on the smaller NGO(110) grow under large compressive strain (1.8%); therefore, coherence with the substrate required films of less than 20 nm thickness.

The "coherent growth" statement does not rule out the existence of orthorhombic variants, i. e., locally different orthorhombic orientations. A separation of orthorhombic film reflections was not feasible with in-house XRD due to the small expected peak splitting. Orthorhombic single or multi-domain states have been indirectly distinguished by the results of MOKE measurements. Sharp rectangular magnetization loops have been observed and taken as sign of an orthorhombic single domain in the field-of-view (the imaged area of about 0.5 $mm^2$), see section "MOKE vs strain". MOKE microscopy gives direct evidence for magnetic single-domain states in an area of 0.25x0.1 $mm^2$. In contrast, films with rounded hysteresis loops have been assigned to contain multiple orthorhombic domains (Fig.2, right panel). A zero longitudinal MOKE signal in two orthogonal in-plane directions indicates an orthorhombic single domain with out-of-plane ***c***-axis <u>or</u> a multi-domain state with cancellation of domain signals.

## Magnetization of a bulk $LaFeO_3$ crystal

A platelet has been cut from the grown crystal with its normal along a pseudocubic direction. It contains orthorhombic domains with well-aligned pseudocubic directions. Reciprocal space maps of x-ray reflections have been taken to probe the multi-domain structure (Supplement, Fig.S6). The orthorhombic domains can be distinguished in the crystal, even though the distortions with respect to the cubic structure are very low.

Field-dependent magnetization loops taken at 300 K and 10 K are shown in Fig.5. They essentially agree with data from literature reports (*9*, *22*). The remanent magnetization perpendicular to the platelet is recorded as 0.041 $\mu_B$/Fe (0.048 $\mu_B$/Fe) at 300 K (10 K). This value is within 10% of the remanence along the ***c***-axis of 0.044 $\mu_B$/Fe at 300 K reported by Treves et al. (*22*). The remanent magnetization measured along an arbitrary in-plane direction is very small, approximately 0.003 $\mu_B$/Fe at 300 K. According to published work on untwinned crystals (*22*), the other orthorhombic axes (***a***, ***b***) show no remanence. Therefore, the remanence values seem to indicate that a large volume fraction (about 90%) of the measured crystal platelet is ***c***-oriented (i. e., the ***c***-axis lies perpendicular to its surface).

The linear susceptibility at 300 K in perpendicular direction is $3.3\cdot10^{-3}$ $\mu_B$/Fe $T^{-1}$. In the in-plane direction, the averaged (not constant) susceptibility is $4\cdot10^{-3}$ $\mu_B$/Fe $T^{-1}$. These values agree reasonably with published susceptibilities along the orthorhombic directions of $3.7\cdot10^{-3}$ $\mu_B$/Fe $T^{-1}$ (***c***-axis), $4.4\cdot10^{-3}$ $\mu_B$/Fe $T^{-1}$ (***b***-axis), and $4.4\cdot10^{-3}$ $\mu_B$/Fe $T^{-1}$ (***a***-axis) reported by Treves et al. (*22*, *23*), supporting the above derived dominating ***c***-orientation of the platelet.

## MOKE measurement of magnetic hysteresis

We measured the two orthogonal in-plane directions parallel to the substrate edges, i. e., approximately the [1$\underline{1}$0] and [001] orthorhombic orientations of the substrate. All films in a coherent compressive strain state (on DSO(110) or NGO(110) substrates) showed a longitudinal MOKE signal. Observations indicate an in-plane magnetization ***M***//**c**//[001] under compressive strain. Films in a tensile strain state (on GSO(110)) fell into two groups: some show longitudinal MOKE indicating in-plane magnetization, others don´t. As said above, out-of-plane magnetization or signal cancellation in a multi-domain state may be the reason. (First checks using the polar MOKE geometry point to the latter case.)

Field-dependent longitudinal MOKE measurements parallel to the substrate edges of compressively strained films showed a sharp rectangular magnetization loop indicating the ***c*** -axis direction (the axis of the magnetization) along [001] as well as a typical hard-axis behavior with linear field dependence in the orthogonal in-plane direction (Fig.2, left panel). The entire film is likely to have the same orthorhombic orientation, since shifting the sample gave no change of the loop shape, just a variation of the coercive field. More confirmation is derived from MOKE microscopy below. The hysteresis loop in Fig.2 (left panel) has a very small field range of switching and, thus, a rectangular shape. More examples are shown in the Supplement (Fig. S4).

Fig.2 (right panel) shows MOKE hysteresis loops with increasing maximum field for a sample which is expected to contain multiple orthorhombic domains. This film is 70 nm thick and, even though it is coherently grown, is of lower structural quality. The loops have increasing saturation magnetization, remanence and coercivity with increasing maximum field up to 800 mT. This behavior is reminiscent of magnetic hysteresis of a polycrystalline ferromagnet. Therefore, we expect this sample to contain structural domains with different ***M*** orientations.

We note that films as thin as 6 nm still show a distinct loop of the field-dependent MOKE, with reduced signal due to the low film thickness.

In a next step, the longitudinal MOKE with an in-plane rotation of the magnetic field direction has been recorded. The angle ($\alpha$)-dependent coercive field follows the relation $H_C(\alpha) = H_C(\alpha = 0)/\cos(\alpha)$, with $\alpha = 0$ for the field applied along the easy magnetization axis (Fig.3). This behavior is well-known for $RFeO_3$ crystals (*20*, *24*) and has been recently reported for $YFeO_3$ films (Wang 2024). Up to magnetic fields of several Tesla (*22*), only a very small rotation of ***M*** occurs in a magnetic field applied at an angle to the ***c***-axis. Therefore, magnetic switching proceeds by domain wall motion only, and the coercivity in a moderate (< 1 T) rotating field follows the Kondorsky model (*20*). This model means in short that only the field component parallel to the easy axis (***c***-axis) has an effect on switching. Our angle-dependent MOKE hysteresis measurements on films agree with the Kondorsky model (Fig.3) and, thus, films mimic the switching behavior of crystals.

MOKE microscopy of domain processes

For films with in-plane magnetization, Kerr microscopy domain images (0.25 x 0.1 $mm^2$) have been taken during field runs of 0 to $H_{max}$ to - $H_{max}$ with varied maximum fields. An exemplary movy of resulting domain processes is provided in the Supplementary materials. Fig.4 shows snapshots from LFO(15 nm)/DSO(110) film in a run with $\mu_0 H_{max}$ = 220 mT which leaves a needle domain of the opposite magnetization orientation stable. The field-free state at the starting point (A) shows a uniform dark contrast and appears to be domain-free within our optical resolution. Subsequent panels (B-E) reveal micrometer-sized magnetic surface defects which may be growth defects. Near the coercive field, small reversed domains appear abruptly. Often, they nucleate at defects. Domain walls are rather irregularly shaped, attributed to the influence of defects. Within a small field range of about 14 mT, large reversed domains expand parallel and perpendicular to the field direction to reverse the magnetization in most of the probed area. This explains the rectangular shape of MOKE hysteresis loops (Fig.2). Defects can clearly be seen to also act as pinning sites.

Magnetic domains of $RFeO_3$ bulk crystals have been explored in earlier published work, recording magnetooptical Faraday rotation in transmission (for example, *25-28*). The magnetic phase present in the ground state of all orthoferrites with non-magnetic R element is the $\Gamma_4$ phase where ***M***//***c*** and ***L***//***a*** (Fig.1, and Ref.*9*)). (We note that care must be taken with the assignment of ***a*** and ***b*** axes which varies in published work. Both are nearly identical in lattice parameter, Tab.1.) Rossol et al. assume the magnetization is constrained to rotate in the ***ac*** plane (*27*). ***C***-oriented crystal platelets of $TmFeO_3$ (at 300 K, in the $\Gamma_4$ phase) revealed a strong preference for Néel domain walls forming stripes parallel to the ***bc*** crystal plane, whereas Bloch walls would be parallel to the ***ac*** plane.

Assuming bulk-like behavior, the type of domain walls in our images can be hypothesized: a domain wall perpendicular to the magnetization (***M***//***c***) axis is a crystallographic ***ab*** wall and would be a Néel wall. The imaged domain walls parallel to ***M*** could lie in ***ac*** or ***bc*** planes, being Bloch walls or Néel walls, respectively. The rotation plane of ***M*** (the ***ac*** plane) is tilted by 45° to the film plane, assuming the undistorted bulk lattice of LFO. However, bulk-like behavior is an appropriate assumption only for low-strain, not-too-thin films. The limits for bulk-like behavior need to be explored.

**Discussion**

Longitudinal MOKE measurements have been demonstrated as sensitive tool for imaging magnetic domain processes and for tracking magnetic switching in $LaFeO_3$ films. Employing MOKE hysteresis measurements, the reversal of the weak canted magnetic moment by magnetic field has been monitored. As discussed below, this is expected to include a reversal of the antiferromagnetic vector. Compressively strained, coherently grown films on orthorhombic substrates show large MOKE signals and magnetic single-domain remanence. This promises a pathway for using orthoferrite films in optical experiments, as also suggested by Wang et al. (*15*).

The orientation of the orthorhombic *c*-axis is, thus far, assumed to be identical with the axis of the weak magnetization ***M***, in agreement with the bulk crystal magnetic order. Under this assumption, the ***c***-axis orientation has been determined in a simple way by recording the longitudinal MOKE. In agreement with results from density functional calculations by Mao et al. (*16*) and experiments on $EuFeO_3$ films by Choquette (*17*), an average compressive in-plane strain tends to orient the orthorhombic c-axis along one of the pseudocubic in-plane directions, whereas tensile strain tends to orient the ***c***-axis in the out-of-plane direction. However, there is a second structural effect on the orthorhombic film orientation resulting from coherent transfer of the octahedral rotations from substrate to the film (*17*). This second mechanism favors the parallel alignment of orthorhombic ***c***-axes of film and substrate due to their matching in-phase rotations. For compressive strain on orthorhombic substrates, both mechanisms favor the in-plane alignment of the c-axes of substrate and film. For an average tensile strain, the two mechanisms compete and can give rise to either an in-plane or an out-of-plane c-axis. Some of our tensile-strain films on $GdScO_3$ substrates showed a longitudinal MOKE signal indicating an in-plane ***c***-axis, others did not. In the latter case, polar MOKE has been tried to identify a perpendicular ***c***-axis. No clear case of perpendicular magnetization has been found yet; the absence of a substantial polar MOKE rather indicates a cancellation due to multiple small orthorhombic domains.

In agreement with this situation, reports in published work show different orientations of the unit cell under tensile strain. Zhu et al (*29*) investigated $LaFeO_3$ grown in a tensile strain state on $DyScO_3$(110). An orthorhombic monodomain with the ***c***-axis perpendicular to the film plane is found in electron diffraction images using scanning transmission electron microscopy (STEM). The pioneering report by Park et al. (*10*) underlines the competing strain mechanism, since parallel orientations of ***c***-axes of both, $LaFeO_3$ film and $GdScO_3$(110) substrate, have been assigned in their work and the film is under average tensile strain. The difficult direct identification of the orientation of the orthorhombic $LaFeO_3$ unit cell by XRD together with competing structural mechanisms affecting this orientation underlines the advantage of a magnetooptic measurement for finding the magnetization axis (***M***//***c***).

Fascinatingly, coherently strained epitaxial films show a single magnetic domain in field-free state in the complete area imaged by Kerr microscopy. This area of (0.25 x 0.1) $mm^2$ is only limited by the field-of-view in the experiment; one may suppose most of the entire film area to form a single magnetic domain. The remanent magnetization can be reversed between two stable states. A domain-free remanence of such thin magnetic films is not often reported. We suggest it may be related to the very small magnitude of magnetization leading to weak stray fields, with the stray field energy proportional to $\boldsymbol{M}^2$ being four orders of magnitude below that of common ferromagnets. Micron-sized domains have been observed by Kerr microscopy during the switching process in magnetic field. The switching proceeds as (i) domain nucleation and (ii) motion of domain walls parallel or perpendicular to the in-plane magnetic field. Kerr microscopy movies of domain processes during magnetic switching reveal the presence of various in-plane orientations of domain walls. Obviously, the role of defects prevents the appearance of more ordered domain patterns during the magnetization process.

The enormously large mobility of domain walls known for crystals of various orthoferrites (*28*) is associated with very low coercive fields of few mT to even < 1 mT at 300 K. Our fully oriented (twin-free) LFO films showed coercive fields between 40 mT and 1000 mT at 300 K (Supplement). Hence, our films need at least one order of magnitude larger magnetic fields than crystals for switching. In bulk crystals, coercivity was reported to be governed by the nucleation energy of reversed domains (*20*, *25*), while the domain wall pinning by defects has little impact. On the other hand, the very low reported values of coercive fields in crystals have been attributed to remaining reversed domains in saturated state (*18*). In the Kerr microscopy movies, some surface defects can clearly be seen to act as both, nucleation sites and pinning centers (Fig.4). We conclude that our films reveal an important role of crystalline defects for magnetization switching.

The microscopic switching proposed in published work (*9*, *24*) proceeds via 180° reversal of all $Fe^{3+}$ moments when a domain wall passes through. This implies a simultaneous reversal of the antiferromagnetic vector *L*, the difference of the magnetizations of antiferromagnetically coupled sublattices. In other words, the moderate applied magnetic field could be used to deterministically switch the antiferromagnetic vector by 180°. We are not aware of published direct evidence for the proposed ***L*** reversal. Indirect evidence comes from recent work on the spin Hall magnetoresistance (SMR) of $LaFeO_3/SrTiO_3(001)$ (*7*, *8*). As the authors point out, the Néel vector is much more likely as the source of the SMR switching in an applied magnetic field than the weak magnetization ***M***. Similar investigations have been reported on other antiferromagnets like NiO (*30*), which lack weak ferromagnetism. The spin Hall magnetoresistance has been assigned to the antiferromagnetic ***L*** reorientation there. Hence, magnetization switching by magnetic field in $LaFeO_3$ is likely to also switch the antiferromagnetic vector.

In a sufficiently large magnetic field parallel to the antiferromagnetic axis, some orthoferrites $RFeO_3$ show a spin reorientation transition (SRT) which occurs in a few of them also in field-free state at lower temperatures (*9*). While for $LaFeO_3$ crystals no such SRT has been reported to our knowledge, SRT have been known for $YFeO_3$ (*19*, *31*). Wang 2024 discuss a SRT in strained $YFeO_3$ films at large fields (*15*). Our crystal data show a transition reminiscent of the typical gradual SRT when the field is applied in pristine state at 300 K, whereas it appears to be irreversible after removing the field (Fig.5). The magnetic field applied to films in the reported MOKE experiments is clearly too low to drive a SRT.

It seems promising to use the presented methodology (epitaxial film growth on orthorhombic substrates and MOKE measurements) for other orthoferrites $RFeO_3$ in order to prepare single-domain films and exploit their large potential in antiferromagnetic spintronics and magnonics. $RFeO_3$ in bulk form with magnetic R elements have a more complex magnetic phase diagram in the field-temperature plane, with antiferromagnetic $\Gamma_4$, $\Gamma_2$ and $\Gamma_1$ phases (for an overview see Ref. 9, for $\Gamma$ phases Ref. 32). Orthorhombic twin-free growth is a precondition to measure or utilize the anisotropic magnetic properties. $YFeO_3$, $EuFeO_3$ and $LuFeO_3$ order in the same magnetic $\Gamma_4$ phase like $LaFeO_3$ in low fields at all temperatures. As mentioned, orthorhombic single-domain growth of $EuFeO_3$ and $LaFeO_3$ has been achieved on $GdScO_3(110)$ (*10*, *17*), and for $YFeO_3$ on $NdGaO_3(110)$ (*15*). The magneto-optic tracking of the magnetization easy axis can allow researchers to identify film orientations in growth studies more easily. An open question in this regard is a possible strain-induced change of the magnetic $\Gamma$ phase as addressed theoretically by Mao et al for $LaFeO_3$ on isotropic substrates (*16*). In that work, strong tensile strain around 2 % drives a $\Gamma_4$ to $\Gamma_2$ transition, changing the magnetization axis from crystallographic ***c*** to ***a*** axis.

Finally, we comment on the origin of the volume expansion of most of our $LaFeO_3$ films. It is likely to originate from a small off-stoichiometry. Scafetta et al. explored the impact of cation ratios of Fe:La > 1 or < 1 in films coherently grown on $SrTiO_3(001)$ by molecular beam epitaxy (*33*). They found that the unit cell volume expands in particular for the case of Fe vacancies, whereas oxygen vacancies did not have the strong volume-expanding effect known for many oxide perovskite materials.

## Summary

We have demonstrated that longitudinal magneto-optical Kerr effect measurements provide a powerful and versatile approach to investigate magnetic switching and domain processes in strained $LaFeO_3$ thin films. By growing coherently strained films on orthorhombic scandate and gallate substrates, we achieved strain-controlled single orthorhombic orientations and identified the direction of the weak ferromagnetic magnetization through the presence of a longitudinal MOKE signal.

Compressively strained films exhibit a pronounced Kerr response, rectangular hysteresis loops, and magnetic single-domain remanence over areas exceeding 0.5 $mm^2$, indicating single-crystal-like magnetic behavior. Angle-dependent hysteresis measurements follow the Kondorsky model, confirming that magnetization reversal is governed by domain-wall motion, closely mimicking the behavior of bulk orthoferrite crystals. Kerr microscopy reveals that magnetic switching proceeds via abrupt domain nucleation and rapid domain-wall motion, with structural defects acting as dominant pinning centers and giving rise to coercive fields significantly larger than those of bulk crystals.

The presented methodology enables straightforward identification of magnetic easy axes and domain configurations and can be readily extended to other orthoferrites $RFeO_3$. Although the Néel vector is not directly probed in the present experiments, the observed reversal of the weak ferromagnetic moment is expected to be accompanied by a deterministic 180° switching of the antiferromagnetic vector. This work thus provides an important experimental foundation for exploiting strain-engineered orthoferrite films in antiferromagnetic spintronics, ultrafast magneto-optical studies, and magnonic applications.

## Acknowledgments

This work has been funded by Deutsche Forschungsgemeinschaft, TRR 227 *Ultrafast spin dynamics* (projects A5, A10, B1). We acknowledge A. Stempel for building the angle-dependent MOKE setup.

| | ***a*** (Å) | ***b*** (Å) | ***c*** (Å) | Ref. | $\boldsymbol{a_{pc}}$ (Å) |
|---|---|---|---|---|---|
| $LaFeO_3$ | 5.5563(4) | 5.5654(3) | 7.8542(4) | *34* | 3.930 ($\boldsymbol{a_{bulk}}$) |
| $NdGaO_3$ | 5.43 | 5.50 | 7.71 | *Crystec* data sheet | 3.860 |
| $DyScO_3$ | 5.44 | 5.71 | 7.89 | *Crystec* data sheet | 3.944 |
| $GdScO_3$ | 5.45 | 5.75 | 7.93 | *Crystec* data sheet | 3.963 |

**Tab.1**: Orthorhombic lattice parameters of all used materials (bulk values). The substrate lattice parameters are given with three digits on the data sheet provided by the company. For $LaFeO_3$, we noted a variation of values reported in published work. The distortions of the cubic structure are very small, (see, e. g., the similarity of ***a*** and ***b***). Therefore, this may be a source of confusion. We decided to use the data of Dixon et al., since these originate from thorough temperature-dependent measurements taken by powder neutron diffraction.

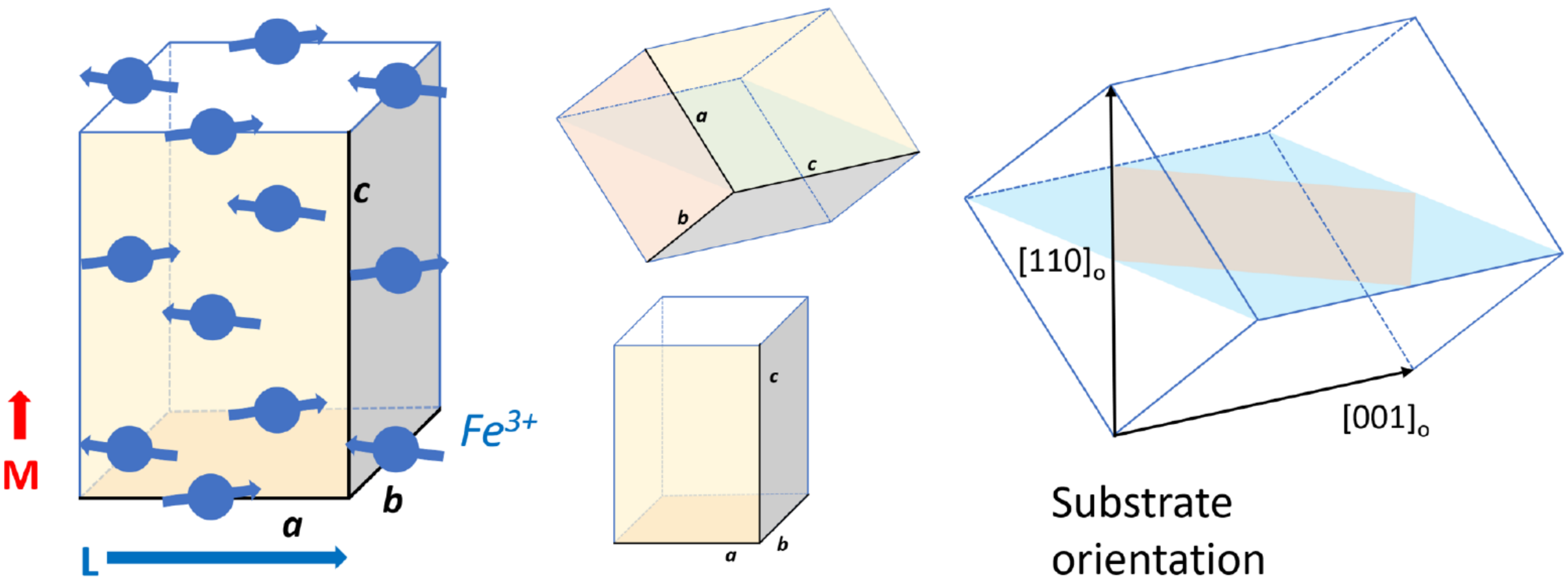

**Fig.1**: Orthorhombic unit cell of bulk $LaFeO_3$ with canted Fe spins and resulting orientations of the weak magnetization (***M***) and the antiferromagnetic vector (***L***) (left panel). Orientation of the substrate (right panel) and the two possible orientations of the orthorhombic unit cell of the film (middle panel).

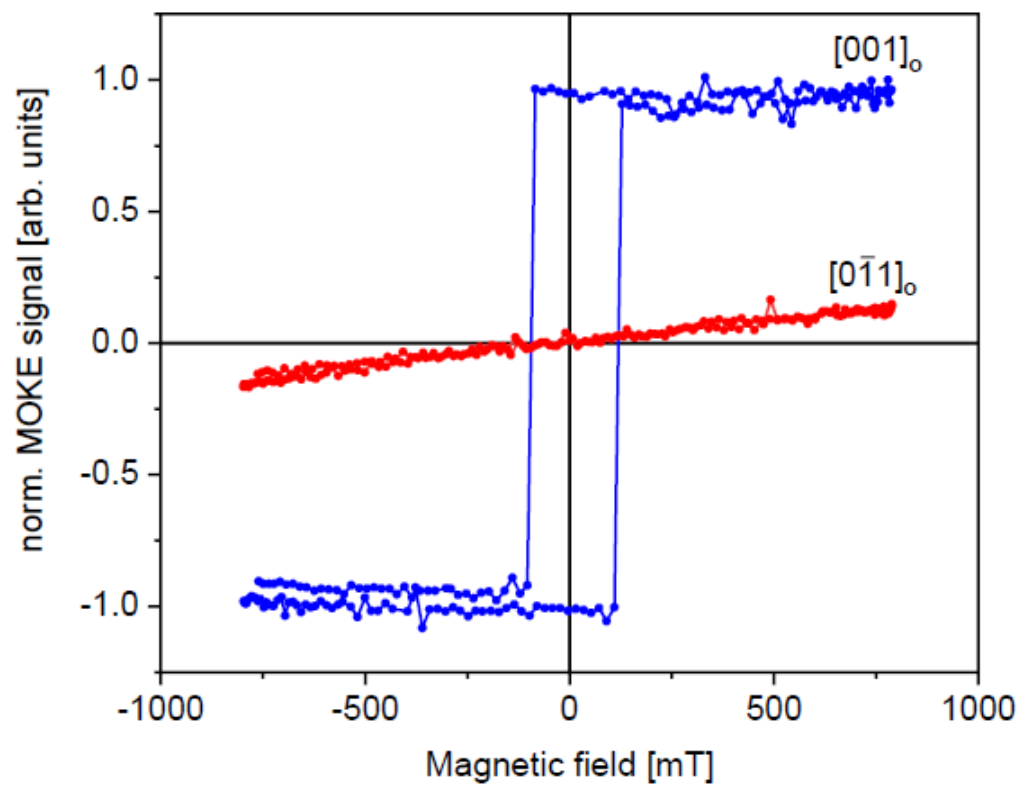

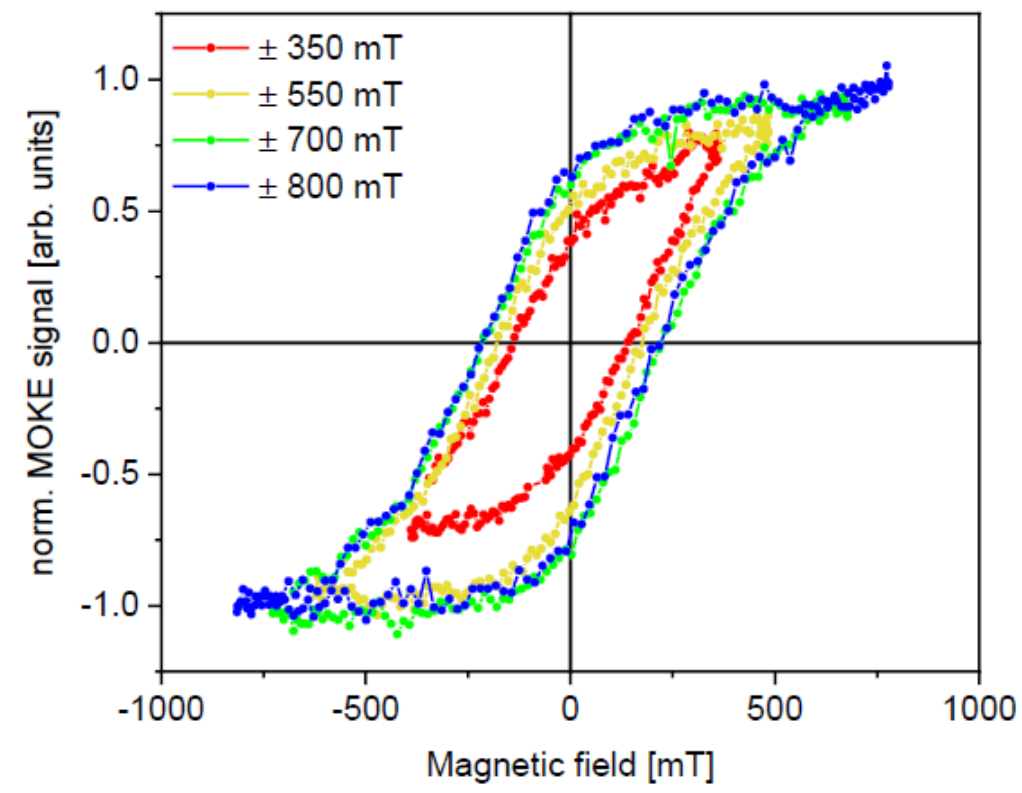

**Fig.2**: Hysteresis loops of longitudinal magneto-optic Kerr effect (MOKE). Left panel: compressively strained $LaFeO_3$(15 nm)/$DyScO_3$(110) film measured along the orthorhombic in-plane directions [001] and [1$\underline{1}$0]. Right panel: hysteresis loops with increasing maximum field of a $LaFeO_3$(69 nm)/$DyScO_3$(110) film. The loop shape indicates multiple magnetic domains.

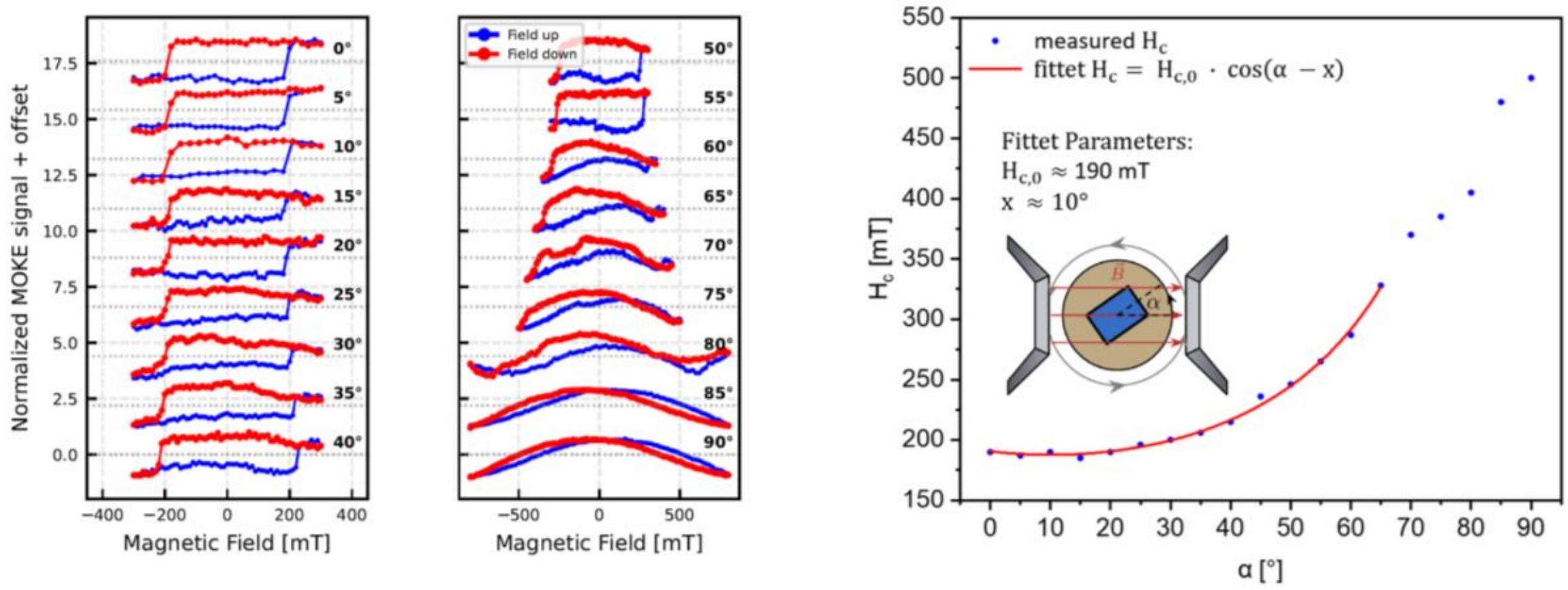

**Fig.3**: Longitudinal MOKE hysteresis loops vs angle ($\alpha$) between the magnetic field and the magnetization axis for a $LaFeO_3$(15 nm)/$DyScO_3$(110) film (left panel). For angles above 65°, loop shapes are distorted, switching cannot be correctly derived. Coercive field vs angle $\alpha$ and fit to the Kondorsky model (right panel). The inset shows the measurement geometry.

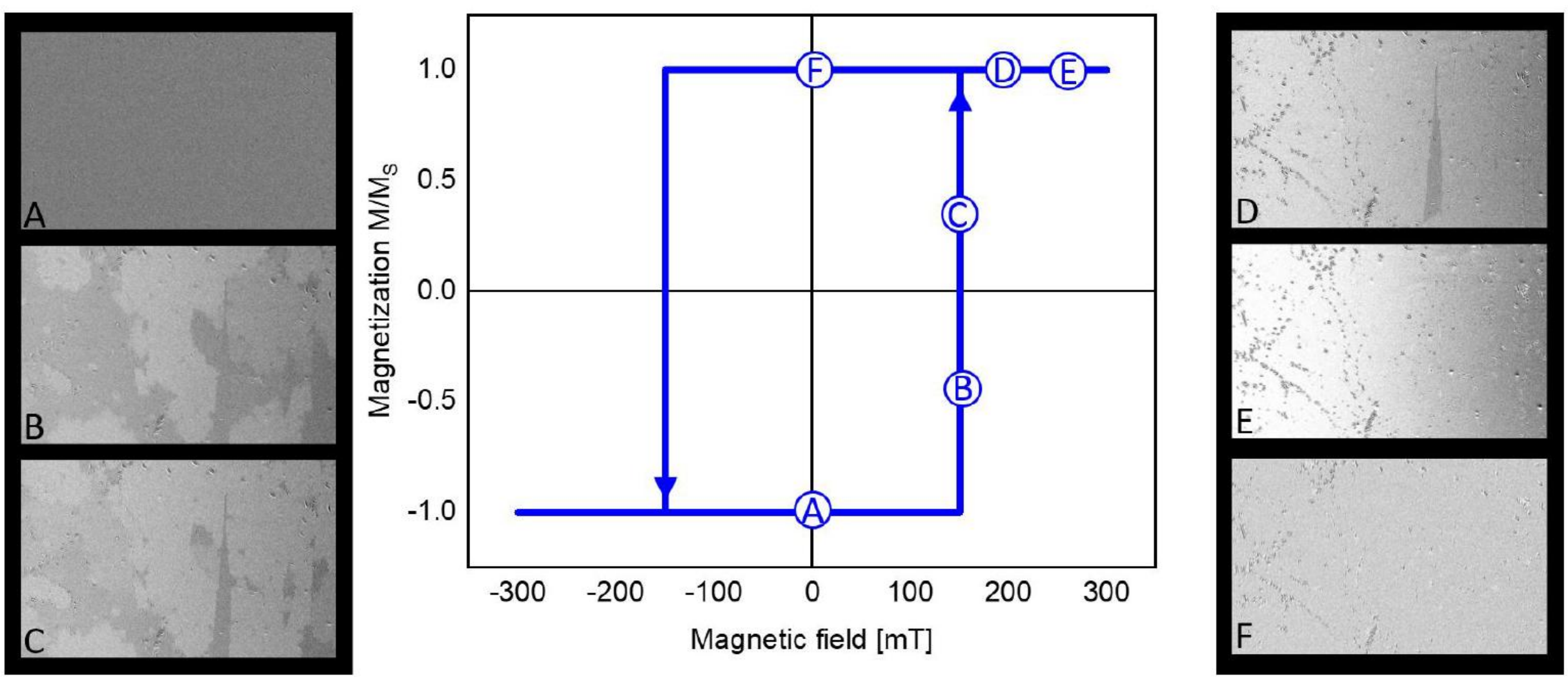

**Fig.4:** Magnetic domains in $LaFeO_3$(15 nm)/$DyScO_3$(110) recorded with Kerr microscopy (longitudinal MOKE) during a magnetic field run. The field-of-view is 0.25 x 0.1 $mm^2$ in size.

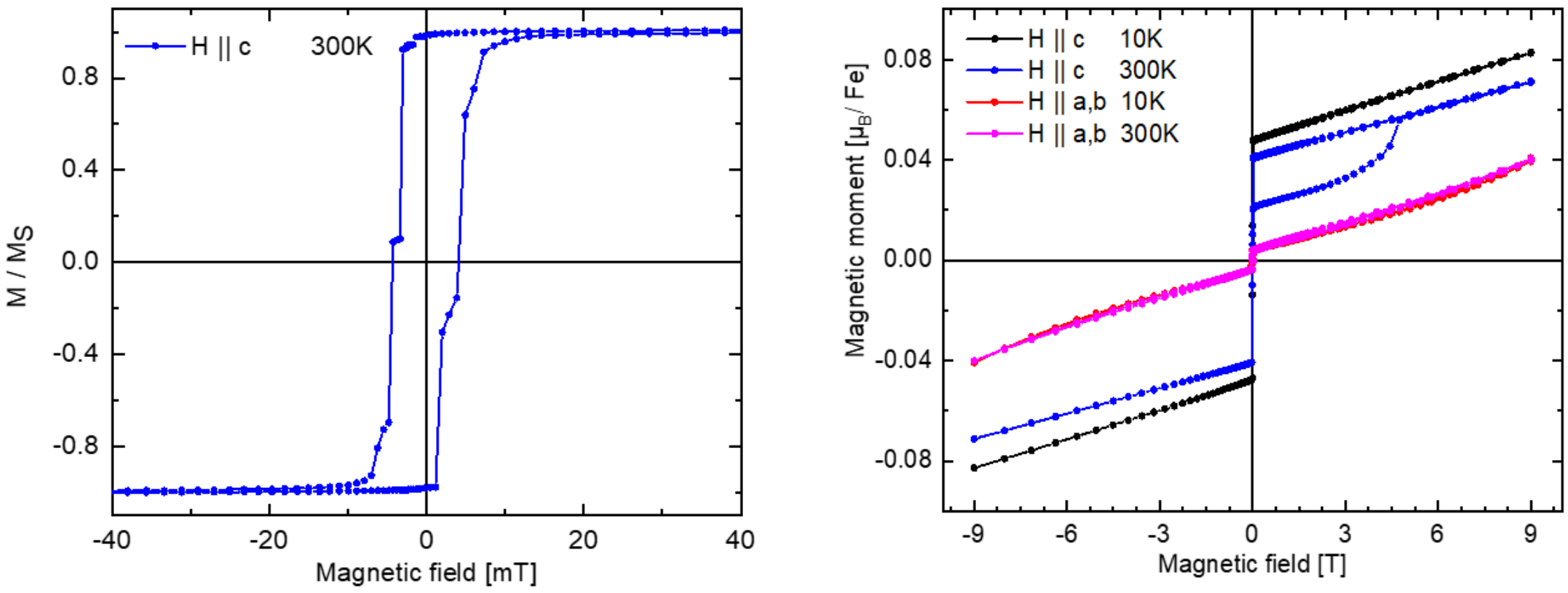

**Fig.5:** Magnetization hysteresis of a twinned $LaFeO_3$ crystal platelet with dominating ***c***-orientation. Left panel: normalized magnetization in the low-field range. Right panel: magnetic moment per Fe atom in fields up to 9 T. The field direction called "***a,b***" lies in the plane of the platelet with deliberate orientation. The transition near 5 T is in pristine state of the sample. Repeated hysteresis measurements did not show this feature which is reminiscent of a field-induced spin reorientation.

# Supplementary Information

## Magneto-optical evidence for single-crystal-like magnetic switching of epitaxial antiferromagnetic $LaFeO_3$ films

A. Rieche,[1] W. Hoppe,[1] C. Körner,[1] A. D. Rata,[1] F. Weber,[2,3] E. M. Vocks,[1] F. Wührl,[1] M. Bargheer,[2,3] W. Widdra,[1] G. Woltersdorf,[1] S. Ebbinghaus,[1] A. Herklotz,[1] K. Dörr[1]

[1]Institute of Physics, Martin-Luther-Universität Halle-Wittenberg, 06099 Halle, Germany

[2]Helmholtz-Zentrum Berlin für Materialien und Energie GmbH, Wilhelm-Conrad-Röntgen Campus, BESSY II, 12489 Berlin, Germany

[3]Institute of Physics and Astronomy, University of Potsdam, Karl–Liebknecht–Str. 24–25, 14476 Potsdam, Germany

**Table of representative samples**

| Substrate | Thickness (nm) | $c_{pc}$ (Å) | Volume increase (%) | L-MOKE | Isotropic in-plane strain (ε, %) |
|---|---|---|---|---|---|
| **$NdGaO_3(110)_o$** | 9 | 3.990 | < 0 | yes | Compressive (-3.4) |
| **$DyScO_3(110)_o$** | 8 | 3.975 | 2.1 | yes | Compressive (-0.75) |
| | 47 | 3.965 | 1.9 | yes | Compressive (-0.5) |
| | 15 | 3.952 | 1.5 | yes | Compressive (-0.2) |
| | 6 | 3.946 | 1.2 | yes | Low strain (-0.05) |
| | 36 | 3.946 | 1.2 | no | Low strain (-0.05) |
| | 70 | (#) | (#) | yes | Multi-domain (#) |
| **$GdScO_3(110)_o$** | 59 | 3.960 | 2.4 | no | Low strain (0.08) |
| | 26 | 3.930 | 1.7 | yes | Tensile (0.3) |

**Table S1.** $LaFeO_3$ films have been coherently grown by pulsed laser deposition on orthorhombic substrates. Coherent growth has been checked recording X-ray reciprocal space maps (like those shown below). Note that "coherence" refers here to the pseudocubic unit cell, since orthorhombic splitting of film peaks could not be resolved. The unit cell volume of the $LaFeO_3$ films is slightly enlarged for most samples, this is listed as relative volume increase. As a consequence, compressive or negligible average film strain has been found on $DyScO_3(110)$ substrates. The average (isotropic) in-plane strain ($\varepsilon = 1 - c_{pc}/a_{pc}$) is calculated as explained in the main text at the beginning of the Results Section. Films thicker than 10 nm on $NdGaO_3(110)$ were not coherently grown. On $DyScO_3$ and $GdScO_3$ substrates, films have been grown coherently up to a thickness of 70 nm. (#) "Multi-domain" means the coexistence of different orthorhombic variants which is concluded from magnetic behavior. The multi-domain film had very broad x-ray reflections; therefore, no value of $c_{pc}$ has been derived.

## Crystalline structure of LFO films and longitudinal MOKE hysteresis curves

### LaFeO$_3$ films on NdGaO$_3$(110)

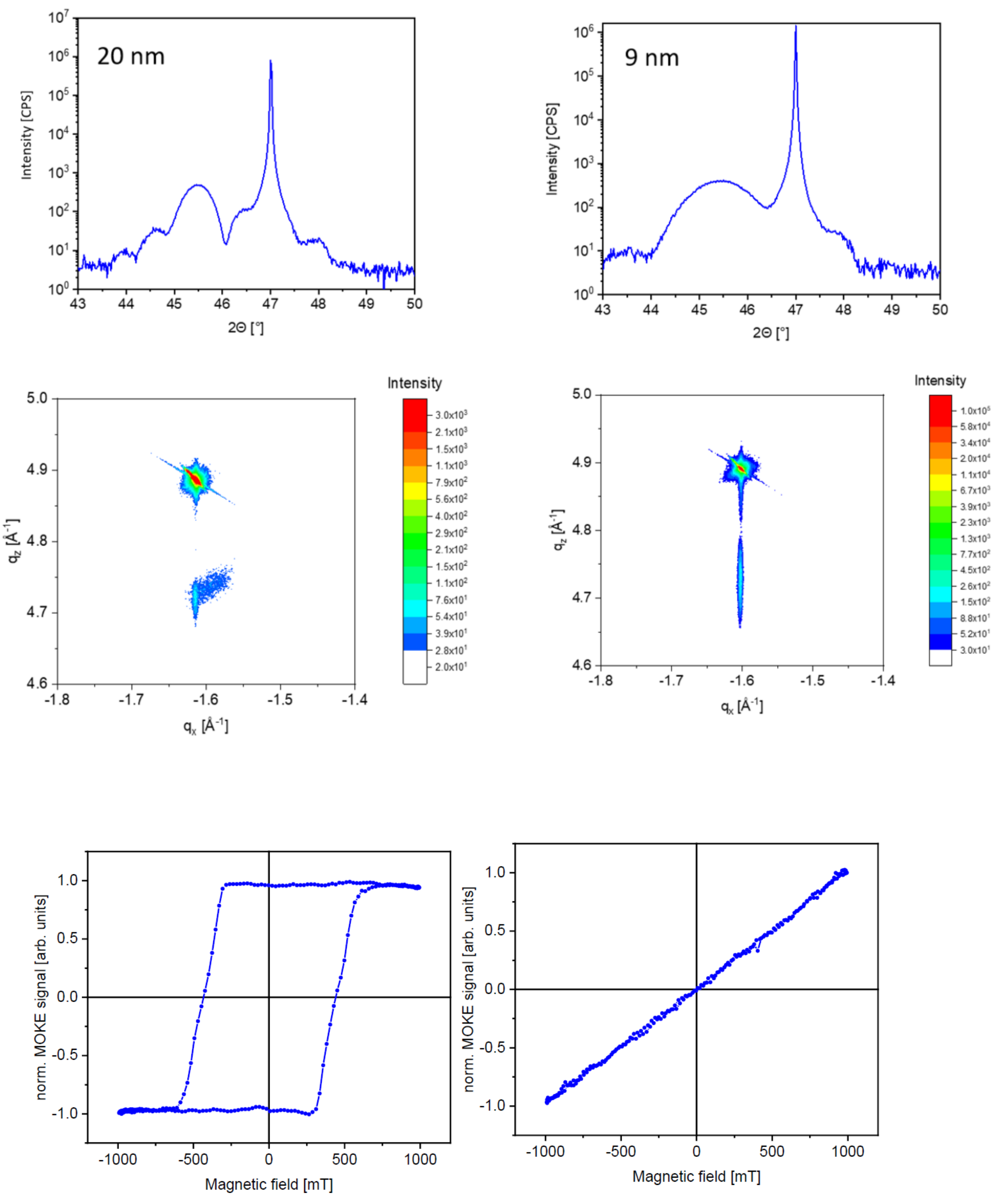

**Fig. S1.** Examples for a coherent (9 nm) and a partially relaxed (20 nm) film on NdGaO$_3$(110). X-ray diffraction $\theta$-2$\theta$ measurements (upper panel) and reciprocal space maps of a $(103)_{pc}$ reflection (middle panel). The lowest panel shows the L-MOKE hysteresis curves of the coherent (9 nm) film in the orthorhombic in-plane directions (left: [001] and right: $[1\bar{1}0]$) of the substrate. The partially relaxed film did not show a L-MOKE signal in any in-plane direction.

## $LaFeO_3$ films on $DyScO_3(110)$

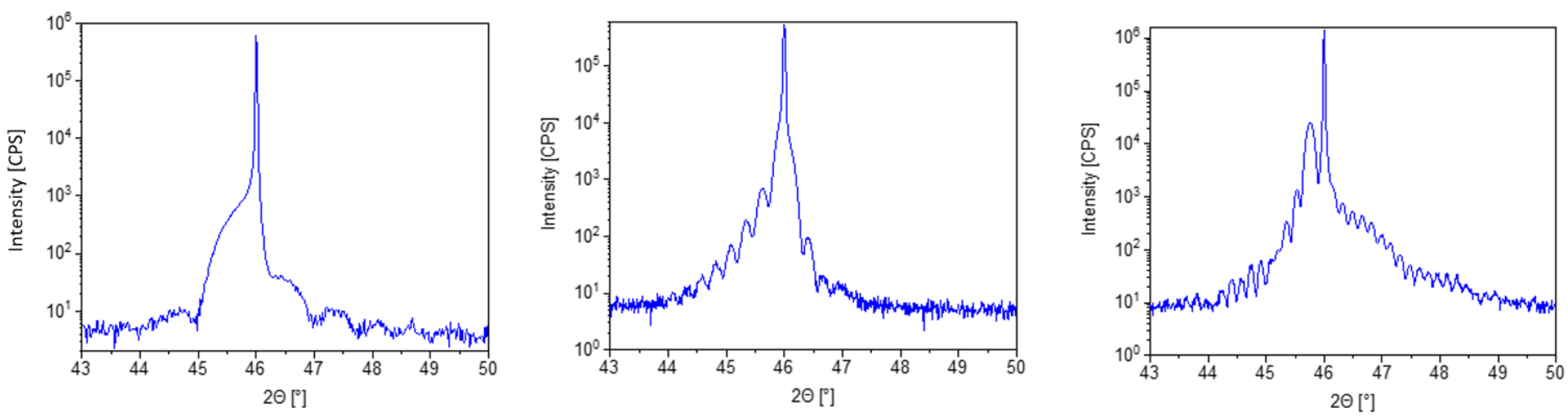

**Fig.S2.** X-ray diffraction θ-2θ measurements of the $(002)_{pc}$ reflection of coherently strained films on $DyScO_3(110)$. Film thicknesses are 18 nm, 30 nm, 40 nm (from left to right). Depending on slight increase of the unit cell volume, films on $DyScO_3(110)$ are compressively strained of nearly strain-free (like the example in the middle panel).

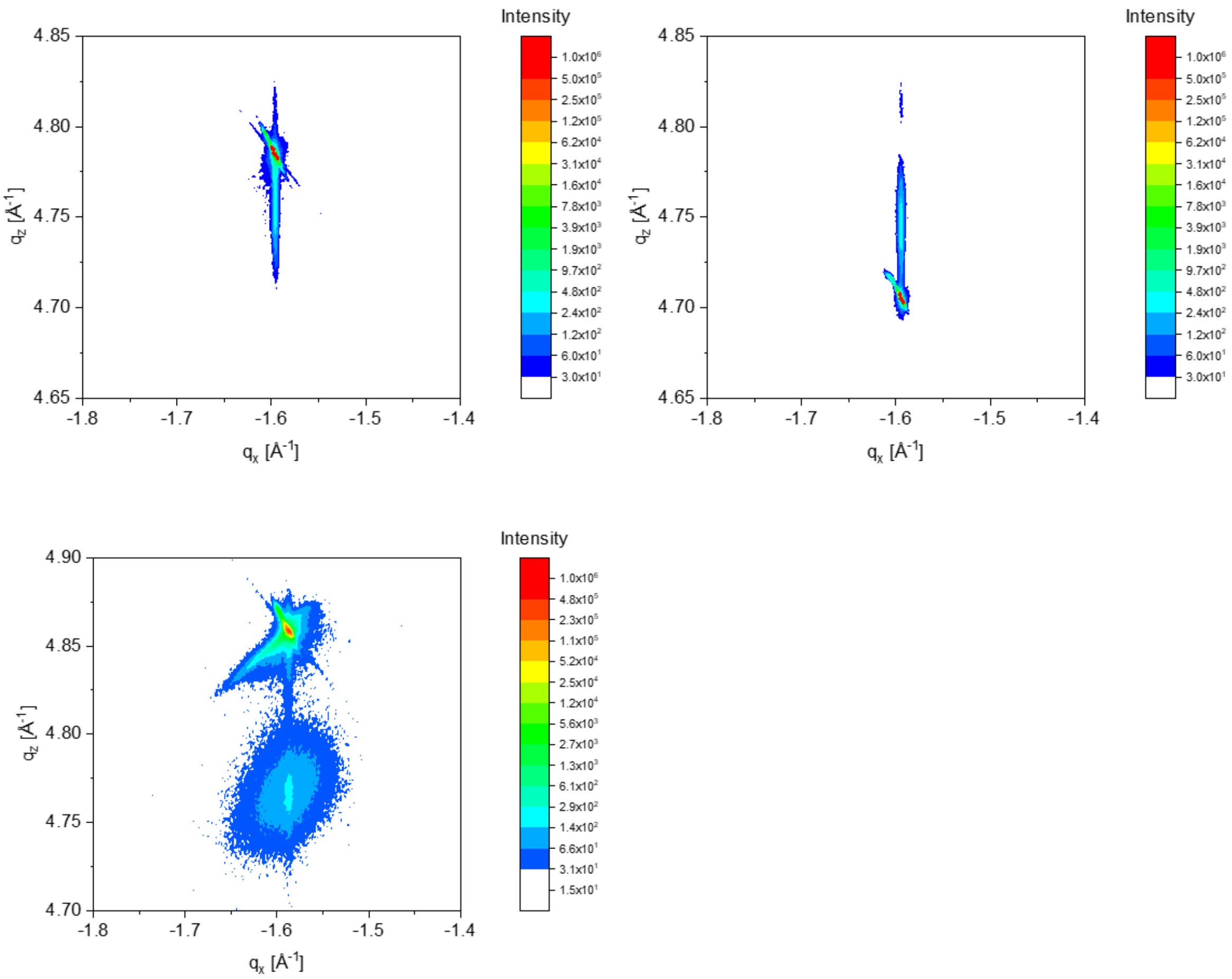

**Fig.S3**. XRD-RSM of $(103)_{pc}$ reflections reveal coherent growth (identical $q_x$ value of film and substrate). Upper panel: an in-plane rotation of the sample by 90° allows for measurement of two different orthorhombic substrate reflections belonging to the pseudocubic $(103)_{pc}$ group. Two different substrate $q_z$ positions indicate the in-plane anisotropy of the substrate. The film peak stays at the same $q_z$ position because of the width of the film peak and the small deviation of $LaFeO_3$ from cubic structure. This is a film with very low average strain; the strain appears to be tensile in one in-plane direction and compressive in the other. Lower panel: XRD-RSM of an $(103)_{pc}$ reflection of the multi-domain film (70 nm) on $DyScO_3(110)$ from Tab.S1. Only one broad film reflection is visible, orthorhombic variants cannot be resolved.

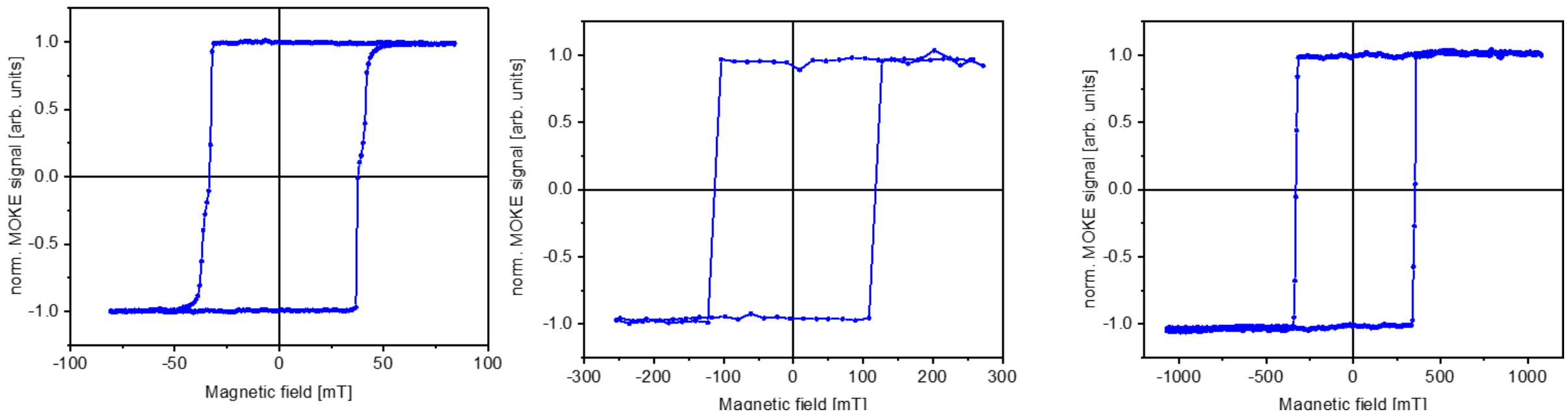

**Fig.S4.** Rectangular hysteresis curves measured by Longitudinal MOKE along the easy axis of magnetization (***M***) for three $LaFeO_3/DyScO_3(110)$ films. Film thicknesses are from left to right: 9 nm, 15 nm, 20 nm. Following the bulk crystal behavior, the easy axis of ***M*** is parallel to the orthorhombic ***c***-axis (or $[001]_o$ axis) of the film which is parallel to the $[001]_o$ axis of the substrate. All samples have also been checked along the orthogonal in-plane direction and show hard-axis behavior.

Note that we are not able to exclude an out-of-plane canting of the magnetization, based on measurements of longitudinal MOKE alone. However, the low-strain films on $DyScO_3(110)$ substrates are likely to follow the bulk behavior, provided one can neglect interface effects on their magnetic order. Substantial strain might reorient the ***M*** easy axis in the crystallographic lattice. Whether or not this happens is yet an open question.

The three examples of hysteresis curves above differ in their coercive fields by an order of magnitude. Several parameters may affect the value of coercive fields, including film thickness und epitaxial strain. Kerr microscopy movies (data are available upon request) of domain switching imply an important role of defects. Therefore, we cannot derive any correlation of coercivity with strain or film thickness in present stage. Clearly, the films have 1-2 orders of magnitude larger coercivity than the measured single crystal (of about 5 mT).

## $LaFeO_3$ films on $GdScO_3(110)$

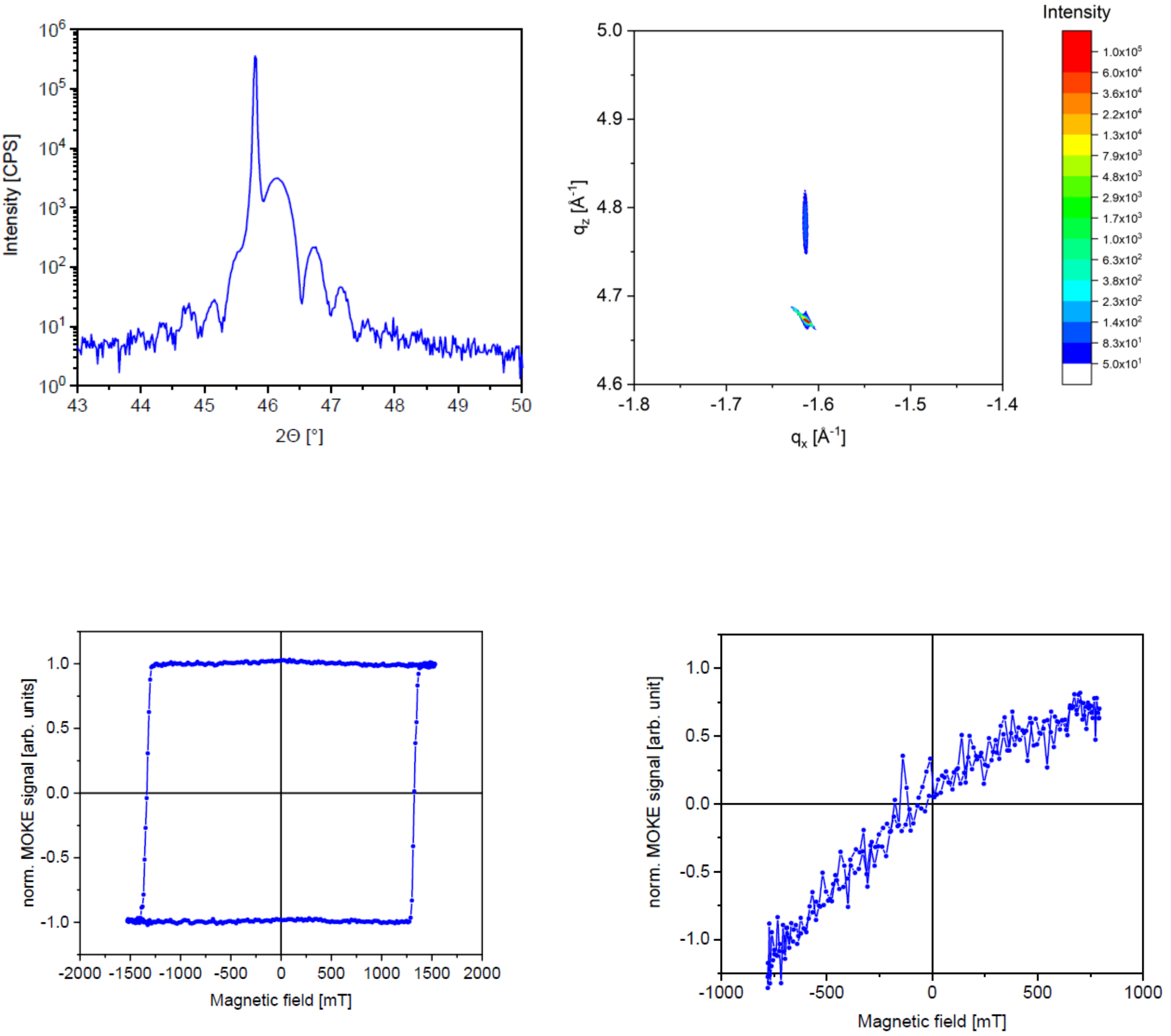

**Fig.S5**. X-ray diffraction θ-2θ measurement and reciprocal space map of a $(103)_{pc}$ reflection of a coherently strained film (26 nm) on $GdScO_3(110)$ (upper panel). This film reveals a rectangular longitudinal MOKE hysteresis loop with in-plane easy axis along the $[001]_o$ axis of the substrate (lower left panel). The orthogonal in-plane direction follows hard-axis behavior. A similar, thicker film (Tab.S1) did not show any longitudinal MOKE (example of a measurement in lower right panel). It is likely that the film has cancelled Kerr signals from multiple small magnetic domains. The logical alternative would be a magnetization axis perpendicular to the film plane; the latter should be observable by polar MOKE. (In ongoing work, we found no significant polar MOKE thus for the sample.)

**XRD reciprocal space map of the $LaFeO_3$ crystal**

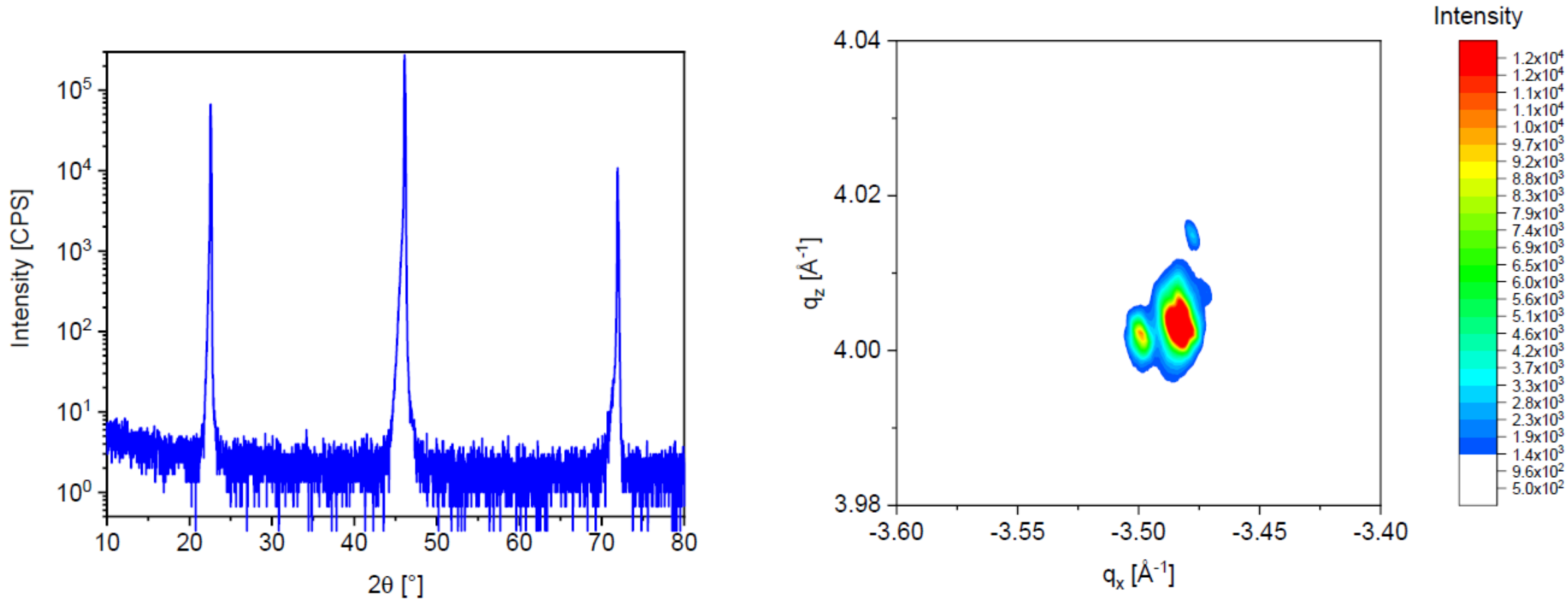

**Fig.S6**. X-ray diffraction θ-2θ measurement of the crystal platelet used for magnetization measurements (left panel). One of the pseudocubic $(113)_{pc}$ reflections in a reciprocal space map (RSM) of the crystal platelet is clearly split into several distinct maxima (right panel). The different orthorhombic domains can account for triple peak splitting of this reflection, provided the pseudocubic orientation is perfect. The apparent larger number of observed maxima is likely to result from small-angle-tilted parts of the crystal.

**Determination of film orientation by Low Energy Electron Diffraction (LEED)**

At present, the orthorhombic orientation of $LaFeO_3$ films is not easily accessible. MOKE measurements identify the magnetization axis which is in bulk crystals parallel to the ***c***-axis (***M***//***c***). However, epitaxial strain might (in principle) change this relation. Low-energy electron diffraction (LEED) of the LFO film surface is a useful tool for the direct identification of the orthorhombic orientation (*1*). This technique requires a clean film surface (using *in-situ* LEED in the growth chamber, or transferring the sample in a vacuum suitcase). In addition, samples need to be sufficiently conducting to avoid charging. The latter is not the case for our LFO films grown directly on the orthorhombic substrates. Therefore, the insertion of a conducting $SrRuO_3$ electrode under the LFO film has been investigated. It was found that the $SrRuO_3$ interlayer can alter the crystallographic orientation of LFO films on DSO(110). Locally different orthorhombic orientations of LFO on $SrRuO_3$/DSO(110) have been identified (*1*). This points to the possibility that $SrRuO_3$ has not grown in a single orthorhombic orientation. In fact, the choice of $SrRuO_3$ as conducting interlayer was not ideal, because $SrRuO_3$ tends to adopt a tetragonal phase for tensile strain around 1 % (*2*, *3*). Vaillionis et al. report on tetragonal $SrRuO_3$ grown on DSO(110) (*2*). Other groups report on orthorhombic $SrRuO_3$ grown on DSO(110) (*4*). We conclude that a $SrRuO_3$ interlayer on DSO(110) is close to a structural phase transition and, thus, can destroy the single-crystalline structural order by forming a mixture of orthorhombic and tetragonal domains.